\documentclass[twocolumn,preprintnumbers,amsmath,amssymb,floatfix]{revtex4}

\usepackage{graphicx}
\usepackage{dcolumn}
\usepackage{bm}

\newcommand{\beq}{\begin{equation}}
\newcommand{\eeq}{\end{equation}}
\newcommand{\bey}{\begin{eqnarray}}
\newcommand{\eey}{\end{eqnarray}}

\begin{document}

\title{Wormhole supported by dark energy admitting conformal motion}

\author{Piyali Bhar}
\email{piyalibhar90@gmail.com} \affiliation{Department of Mathematics, Government General Degree College, Singur,
Hooghly -712 409, West Bengal, India}

\author{Farook Rahaman}
\email{rahaman@associates.iucaa.in} \affiliation{Department of
Mathematics, Jadavpur University, Kolkata 700032, West Bengal,
India}

\author{Tuhina Manna}
\email{tuhinamanna03@gmail.com} \affiliation{Department of Mathematics and Statistics(Commerce Evening), St. Xavier's College,30, Mother Teresa Sarani, Kolkata-700 016, West Bengal, India.}

\author{Ayan Banerjee}
\email{ayan\_7575@yahoo.com} \affiliation{Department of
Mathematics, Jadavpur University, Kolkata 700032, West Bengal,
India}

\date{\today}

\begin{abstract}
In this article, we study the possibility of sustaining a static and spherically
symmetric traversable wormhole geometries admitting conformal motion in Einstein gravity,
which presents a more systematic approach to search a relation between matter and geometry.
In wormhole physics, the presence of exotic matter is a fundamental ingredient and we
show that this exotic source can be dark energy type which support the existence of
wormhole spacetimes. In this work we model a wormhole supported by dark energy which admits conformal motion. We also discuss the possibility of detection of wormholes
in the outer regions of galactic halos by means of gravitational lensing. The studies of
the total gravitational energy for the exotic matter inside a static wormhole configuration are also done.
\end{abstract}


\maketitle

\section{Introduction}
In last two decades, there has been a considerable interest in the field of wormhole physics after
seminal work by Morris-Thorne \cite{Morris}. They proposed the possibility of traversable wormholes
in the theoretical context of the general relativity as a teaching tool. Topologically, wormholes acts as a
tunnels in the geometry of space and time that connect two space-times of same universe or of different universes altogether by a minimal surface called the throat of the wormhole, satisfying  flare-out condition \cite{HV1997},
through which a traveler can freely traverse in both directions. Today, most of the efforts are directed to study the
necessary conditions to ensure their traversability. The most striking of these properties is a special
type of matter that violates the energy conditions, called exotic matter which is necessary to
construct traversable wormholes.
Recent astronomical observations have confirmed that the universe is undergoing
a phase of accelerated expansion which was conformed by the measurements of supernovae
of type Ia (SNe Ia) and the cosmic microwave background anisotropy \cite{Riess}. It has been
suggested that dark energy is still an unknown component with a relativistic negative pressure,
is a possible candidate for the present cosmic expansion and our Universe is composed of approximately 70 percent of it.
The simplest candidate for explaining the dark energy is the cosmological constant $\Lambda$ \cite{Carmelli},
which is usually interpreted physically as a vacuum energy, with $p= -\rho$. Another possible way to explain
the dark energy by invoking an equation of state, p= $\omega\rho$ with $\omega< 0$, where p is the spatially homogeneous pressure and $\rho$ the energy
density of the dark energy, instead of the constant vacuum energy density. As a particular range of the $-1 <$  $\omega<-1/3$, is a widely
accepted results known as quintessence is often considered. The ratio $\omega< -1$ has been denoted phantom energy,
corresponding to violation of the null energy condition, thus providing a theoretically supported scenario
for the existence of wormholes \cite {Sushkov}. The presence of phantom energy in the universe
leads to peculiar properties, such as Big Rip scenario \cite{Caldwell}, the black hole mass decreasing
by phantom energy accretion \cite{Babichev}. Therefore, the dark energy plays an important role in cosmology naturally makes us search
for local astrophysical manifestation of it. In the present work we consider wormhole solution containing dark energy as
equation of state.

The gravitational lensing (GL) is a very useful tool of probing a number of interesting phenomena of the universe.
Particular it can provide rich information for the structure of compact astrophysical
objects like e.g., black holes, exotic matter, super-dense neutron stars, wormholes etc. Out of this, the observation of
Einstein ring and the double or multiple mirror images are the powerful examples for gravitational lensing effect \cite{Hewitt}.
In earlier works GL phenomenon has been studied in the weak field (see \cite{Schneider}), but success leads to explore
other extreme regime, namely, the GL effect in the strong gravitational field has been studied by \cite{Virbhadra}.
Out of various intriguing objects mentioned above, recently it was proposed that wormholes
can act as gravitational lenses and induce a microlensing signature on a background source studied
by Kim and Sung \cite{Kim} and Cramer \emph{ et al.,} \cite{Cramer} and lensing by negative mass wormholes have
been studied by Safonova \emph{ et al.,} \cite{Safonova}. Related with the issue of GL effects on wormholes have been
studied \cite{Nandi}. Recently, the possibility of detection of traversable wormholes in
noncommutative-geometry is studied by Kuhfittig \cite{Peter K. F. Kuhfittig} in the outer regions of galactic
halos by means of gravitational lensing. The possible existence of wormholes in the outer regions of the halo was
discussed in Ref \cite{Rahaman(2014)}, based on the NFW density profile. One of the aims of the current paper
is to study the effect of lensing phenomenon for the wormhole solutions in the presence of exotic matter such as phantom
fields admitting conformal motion.
The present work has been considered in more systematic approach to find the exact solutions
and study the natural relationship between geometry and matter.   For instance, one may
adopt a more systematic approach (see Ref. \cite{Herrera}) by assuming spherical symmetry
and the existence of a non-static conformal symmetry. Suppose that a conformal Killing vector
 $\xi$ is defined on the metric tensor field g defined by the action of the Lie
 infinitesimal operator $\mathcal{L}_{\xi}$, which leads to the following relationship:
\begin{equation}
\mathcal{L}_\xi  g_{ik}=\psi  g_{ik},
\end{equation}
where $\mathcal{L}$ is the Lie derivative operator and $\psi$  is the conformal factor.
Here the vector $\xi$ generates the conformal symmetry in such a way that
the metric g is conformally mapped onto itself along $\xi$.  For an interesting observation
neither $\xi$ nor $\psi$ need to be static even though one considers a static metric.  For $\psi = 0$ then Eq. (1) gives the killing vector, for $\psi = constant$ Eq. (1) gives a homothetic vector and if
$\psi = \psi(x,t)$ then it gives conformal vectors. Further note that when
$\psi = 0$ the underlying spacetime is asymptotically flat which implies that the
Weyl tensor will also vanish. Thus we can develop a more vivid idea about the
spacetime geometry by studying the conformal killing vectors.
 Recently, Bohmer et al. \cite{Bohmer} have studied the traversable wormholes
under the assumption of spherical symmetry and the existence of a non-static conformal symmetry.

The outline of the present paper is as follows: In Sec.
\textbf{II.} we give a brief outline of the conformal killing
vectors for spherically symmetric metric while in Sec.
\textbf{III.} we present the structural equation of phantom energy
traversable wormholes and discuss the physical properties of our solution in the outer region of the
halo by recalling the movement of a test particles. In Sec. \textbf{IV.} we present the stability
of wormholes under the different forces where the total gravitational energy for the exotic matter distribution in the
wormhole discuss in  Sec. \textbf{V.} In Sec. \textbf{VI.} gravitational lensing has been
studied and the angle of surplus are calculated. In Sec. \textbf{VII.} the
interior wormhole geometry is matched with an exterior Schwarzschild solution at the junction interference.
Finally, in Sec. \textbf{VIII.} we discuss some specific comments regarding the results
obtained in the study.

\section{Einstein field equations and conformal killing vector}

The spacetime metric representing a static and spherically symmetric
line eleminent is given by
\begin{equation}
ds^{2}=-e^{\nu(r)}dt^{2}+e^{\lambda(r)}dr^{2}+r^{2}(d\theta^{2}+\sin^{2}\theta d\phi^{2}),
\end{equation}
where $\lambda$ and $\nu$ are functions of the radial coordinate, r.
We shall assume that our source is filled with an anisotropic fluid distribution and
using the Einstein field equation $G_{\mu\nu}= 8\pi T_{\mu\nu}$, for the above metric,
which in our case read (with c = G = 1)
\begin{eqnarray}
e^{-\lambda}\left[\frac{\lambda'}{r}-\frac{1}{r^{2}} \right]+\frac{1}{r^{2}}=8\pi \rho,\\
e^{-\lambda}\left[\frac{1}{r^{2}}+\frac{\nu'}{r} \right]-\frac{1}{r^{2}}=8\pi p_r,\\
\frac{1}{2}e^{-\lambda}\left[ \frac{1}{2}\nu'^{2}+\nu''-\frac{1}{2}\lambda'\nu'+\frac{1}{r}(\nu'-\lambda')\right]=8\pi p_t,
\end{eqnarray}
where $\rho$, $p_r$ and $p_t$ denotes the matter density, radial and transverse pressure respectively of the underlying fluid distribution. and `$\prime$' denotes differentiation with respect to the radial coordinate r.\par
Applying a systematic approach in order to get exact solutions, we demand that the interior spacetime admits conformal motion
 (but neither $\xi$ nor $\psi$ need to
be static even though for a static metric) and therefore Eq. (1) provides the following relationship:
\begin{equation}
\mathcal{L}_\xi g_{ik}=\xi_{i;k}+\xi_{k;i}=\psi g_{ik},
\end{equation}
with $ \xi_{i}=g_{ik}\xi^{k}$. The above equation gives the following set of expressions as
\begin{equation}
\xi^{1}\nu'=\psi,~~~\xi^{4}=C_1,~~~\xi^{1}=\frac{\psi r}{2}~~~and~~~\xi^{1}\lambda'+2\xi^{1},_1=\psi,
\end{equation}
where $C_1$ is a constant and the conformal factor is independent of time i.e., $\psi  =\psi(r)$.
Now, the metric (2), and using the Eq. (6-7) provides the following results:
\begin{eqnarray}
e^{\nu}=C_2^{2}r^{2},\\
e^{\lambda}=\left(\frac{C_3}{\psi}\right)^{2},\\
\xi^{i}=C_1\delta_{4}^{i}+\left( \frac{\psi r}{2}\right)\delta_1^{i},
\end{eqnarray}
where $C_2$ and $C_3$ are constants of integrations.\par
 An important note of this
solutions that immediately ruled out, is that the conformal
factor is zero by taking into account Eq. (9), at the throat of the wormhole
i.e., $\psi(r_0) = 0$, where $r_0$ stands for location of the throat of the wormhole.
Now, using Eqs. $(8)-(10)$, one can obtain the expression for Einstein field equations as
\begin{eqnarray}
\frac{1}{r^{2}}\left[1-\frac{\psi^{2}}{C_3^{2}}\right]-\frac{2\psi \psi'}{r C_3^{2}}=8\pi \rho,\\
\frac{1}{r^{2}}\left[\frac{3\psi^{2}}{C_3^{2}}-1\right]=8\pi p_r,\\
\frac{\psi^{2}}{C_3^{2}r^{2}}+\frac{2\psi \psi'}{r C_3^{2}}=8\pi  p_t.
\end{eqnarray}

Observing the Eqs. $(11)-(13)$, we have three equations with four unknowns namely
$\rho$, $p_r$, $p_t$ and $\psi(r)$ respectively. In order to solve the system of equations,
we need an equation of state relating matter and density by the following simplest relation
$p= p(\rho)$.

\section{Solution for phantom wormhole and physical analysis}

According to Morris and Throne \cite{Morris}, for constructing a wormhole solution one require an unusual
form of matter known as `extotic matter', which is  the fundamental ingredient to sustain traversable wormhole.
The characteristic of such matter is that the energy density $\rho$ may be positive or negative but the
radial pressure $p_r$ must be negative. Theoretical advances shows that the expansion of our
present universe is accelerating and dark energy is a suitable candidate to explain this cosmic expansion.
In this context, we study the construction of traversable wormholes, using the phantom energy equation
of state by the following relationship
\begin{equation}
p_r=\omega \rho~~~~~~~~~~with~~\omega<-1,
\end{equation}
by taking into account Eqs. (11) and (12), with the help of equation (14) we obtain
\begin{equation}
\psi^{2}=\left(\frac{\omega+1}{\omega+3}\right)c_3^{2}+\psi_0r^{-\left(\frac{\omega+3}{\omega}\right)},
\end{equation}
where $\psi_0$ is the constant of integration. For convenience we rewrite the Eqs. (11)-(13), using Eq. (15)
with new dimensionless parameters $\tilde{\psi_0}$=$\frac{\psi_0}{c_3^2}$,
we obtain the expression of matter density, radial and transverse pressure as
\begin{eqnarray}
\rho&=&\frac{1}{8\pi}\left[\frac{2}{r^{2}\left(\omega+3\right)}-\tilde{\psi_0}\frac{\left(2\omega+3\right)}
{\omega}r^{\frac{-3\left(\omega+1\right)}{\omega}}\right],\\
p_r&=&\frac{1}{8\pi}\left[\frac{2\omega}{r^{2}\left(\omega+3\right)}-\tilde{\psi_0}\left(2\omega+3\right)r^{\frac{-3\left(\omega+1\right)}{\omega}}\right],\\
p_t&=&\frac{1}{8\pi}\left[\frac{\omega+1}{r^{2}\left(\omega+3\right)}-\frac{3\tilde{\psi_0}}{\omega}r^{\frac{-3\left(\omega+1\right)}{\omega}}\right].
\end{eqnarray}
Plugging the expression for $\psi^{2}$ given in Eq. (9) with the dimensionless
parameter the expression for metric potential is obtained as
\begin{equation}
e^{-\lambda}=\frac{\omega+1}{\omega+3}+\tilde{\psi_0}r^{-\left(\frac{\omega+3}{\omega}\right)}.
\end{equation}
Thus, taking into account the relation between metric potential and the shape function of the wormhole
by the relation $e^{\lambda}=\frac{1}{1-b(r)/r}$, we obtain form of shape function as
\begin{equation}
b(r)=\frac{2r}{\omega+3}-\tilde{\psi_0}r^{-\frac{3}{\omega}}.
\end{equation}
From the expression of $b(r)$, we see that $\frac{b(r)}{r}$ tends to a finite value as $r\rightarrow \infty $
and the redshift function does not approach zero as $r\rightarrow \infty $, so spacetime is
not asymptotically flat due to the conformal symmetry. \par
Now, we will concentrate to verify whether the obtained
expression for the shape function $b(r)$ satisfies
all the physical requirements to maintain a wormhole solution. For this purpose we are
trying to describe fundamental property of wormholes with help of graphical representation.
The profile of shape function $b(r)$ is plotted in Fig. 1 for the values of
$\tilde{\psi_0}=0.09$ and $\omega=-1.58$, where the flaring out condition has
been checked in Fig. 1 (right panel). We observe that shape function is decreasing
with increase of the radius, and $\frac{db(r)}{dr}<0$ for $r>10.45$. From the left panel
of Fig. 2, we observe that the throat of the wormhole occurs
where $ b(r)-r$ cuts the r axis at a distance $r=5.39$. Therefore the throat of the
wormhole occurs at $r = 5.39$  Km. for our present model. Consequently, we
observe that $b'(5.39) = 0.632 < 1$ and for $r > r_0$ we see that $b(r )-r <0$,
which implies $\frac{b(r)}{r}<1$ for $r > r_0$,
strongly indicate that our solution satisfy all the physical criteria for wormhole solution.
The slope of $b(r)$ is positive upto $r=10.45$, which concludes that the wormhole can not be arbitrarily large.
The same situation occurred in the previous work \cite{bhar}. Moreover, we consider the energy conditions and
the violation of the null energy condition (NEC) i.e., $\rho+p_r <~0$,  is a necessary property for a static
wormhole to exist. In Fig. 2 (right panel), we have studied  all types of  energy condition (using Eqs.(16)-(18)),
graphically and observed that our solution violated the NEC to hold a wormhole open.

\begin{figure*}[thbp]
\begin{center}
\begin{tabular}{rl}
\includegraphics[width=6 cm]{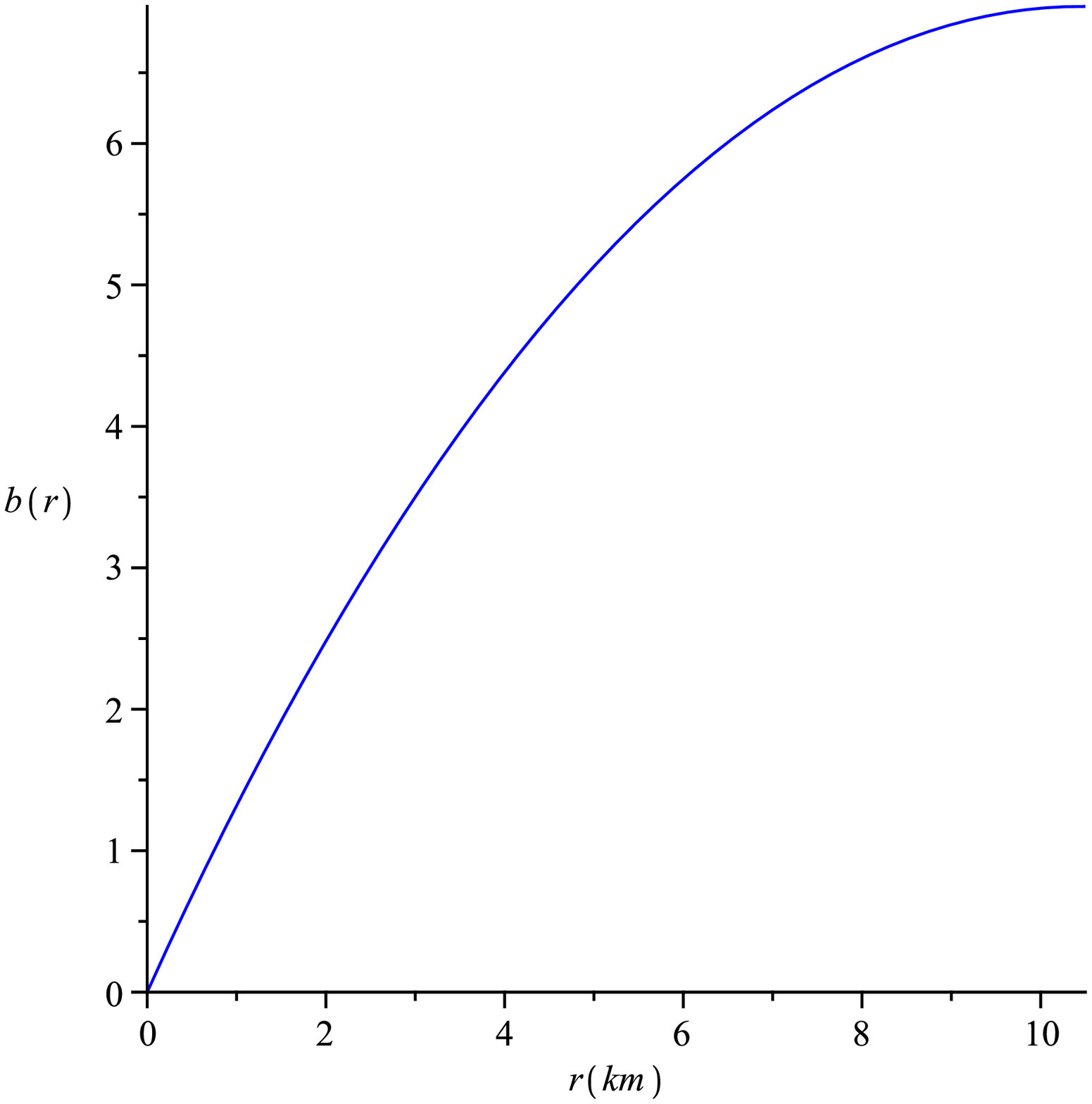}&
\includegraphics[width=6 cm]{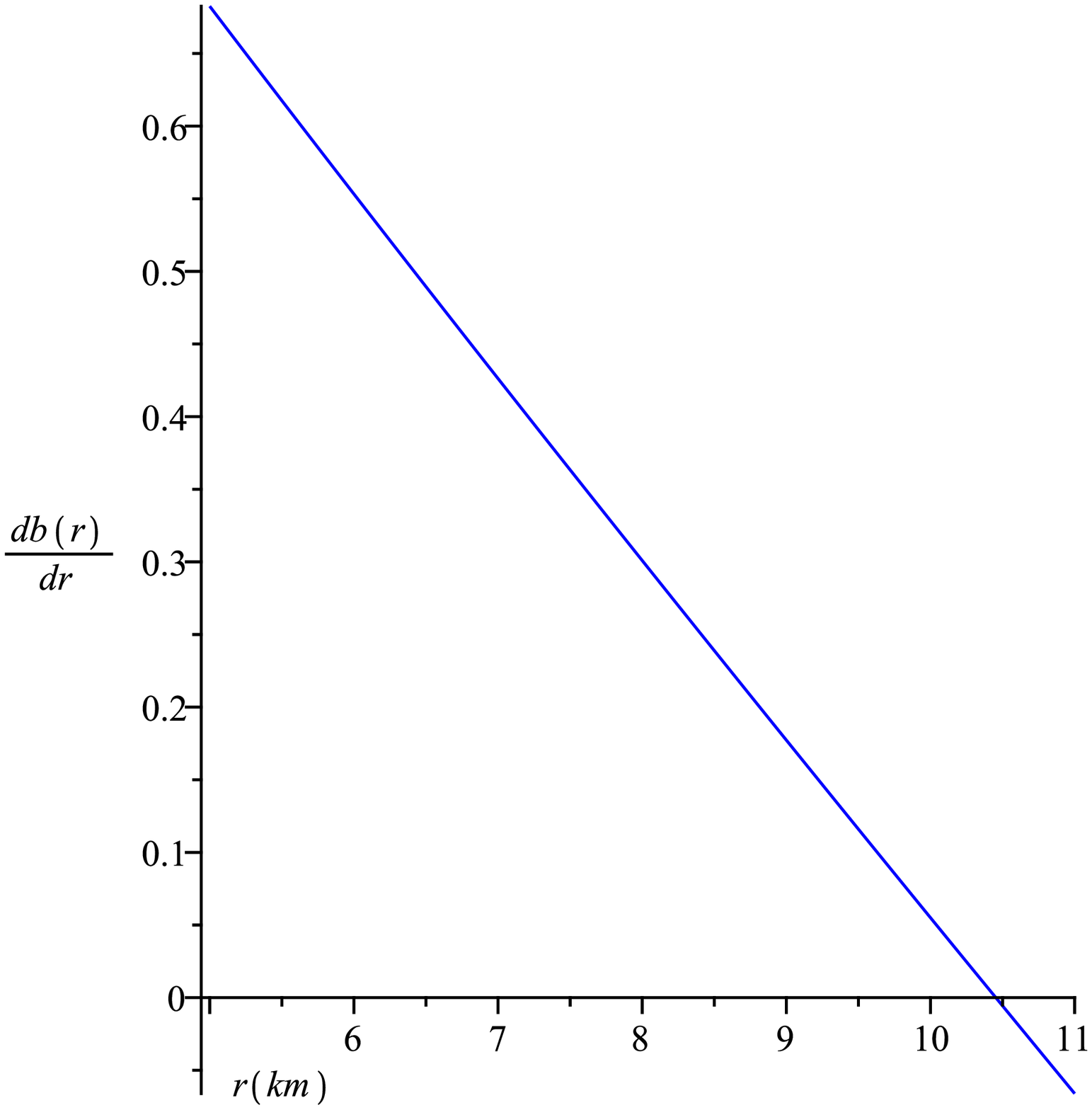}\\
\end{tabular}
\end{center}
\caption{The shape function b(r) is plotted against r {\em left panel}. The values of the parameters are
$\tilde{\psi_0}=0.09$ and $\omega=-1.58$. The slope of shape function of wormhole is plotted
in the {\em right panel} with the same parametric values. The figure indicates that the
slope of $b(r)$ is still positive within the range of r = 10.45.}
\end{figure*}

\begin{figure*}[thbp]
\begin{center}
\begin{tabular}{rl}
\includegraphics[width=6 cm]{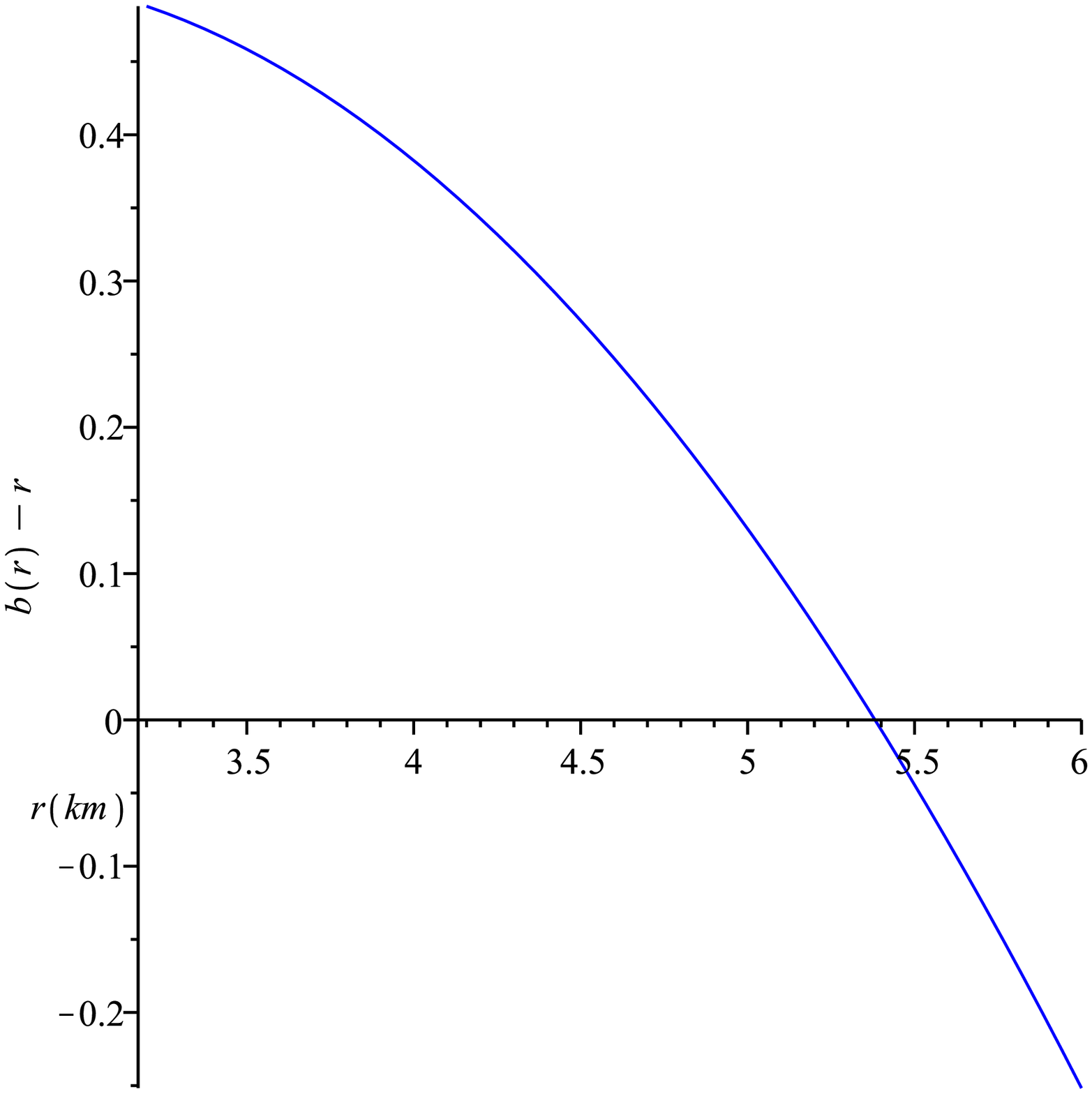}&
\includegraphics[width=6 cm]{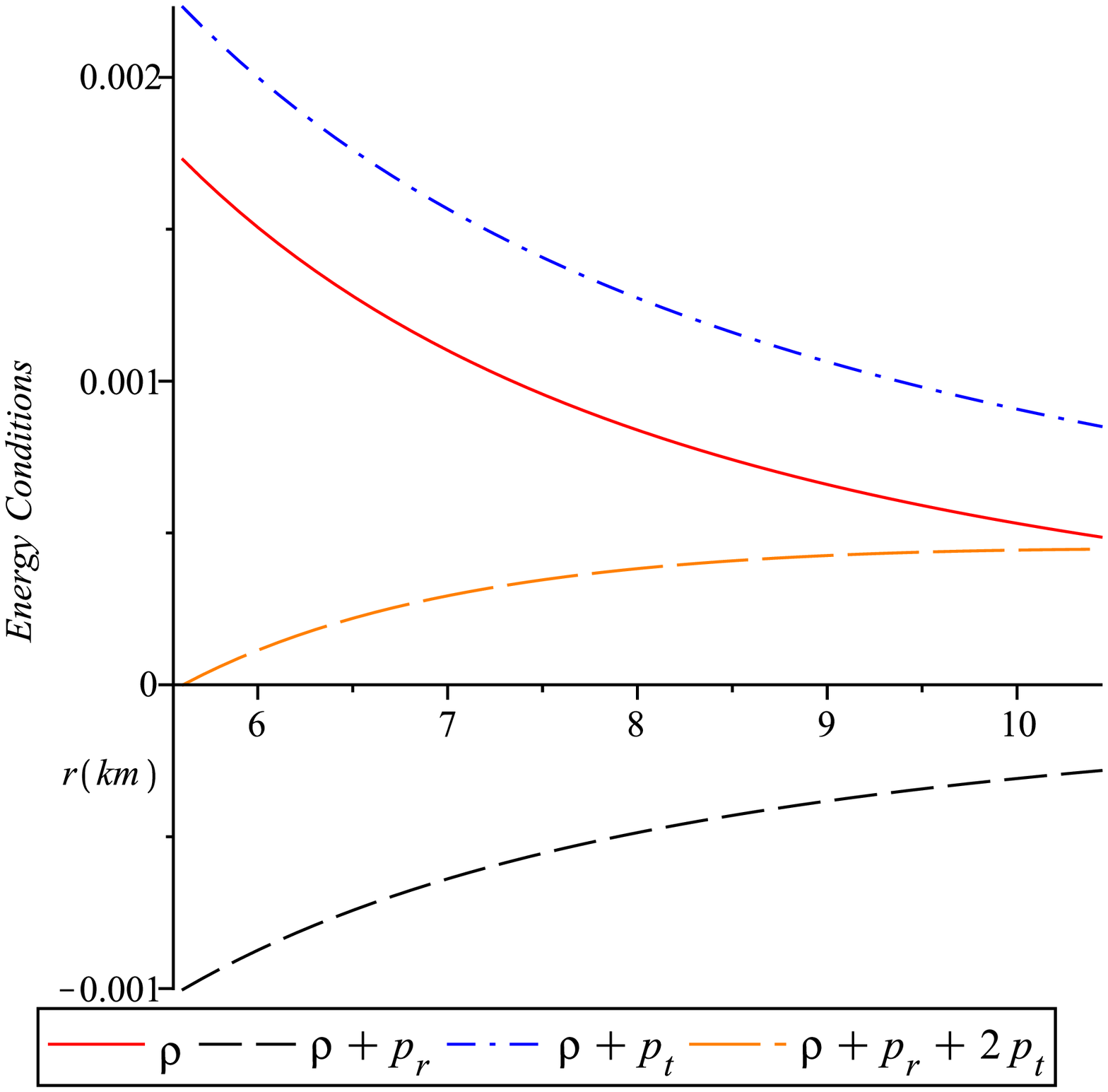}\\
\end{tabular}
\end{center}
\caption{The profile of $b(r)-r$ is plotted against r in the {\em left panel} with the same values of
parameters as stated earlier. The throat of the wormhole occurs where the graph of $b(r)-r$ cuts
the r axis and for our model the throat occurs at r=5.39 Km. The energy conditions are plotted in
the {\em right panel}. The figure indicates the null energy condition is violated for wormhole model.}
\end{figure*}

Since, the wormhole space-time is non-asymptotically flat and
hence the wormhole spacetime should match at some junction radius
$r=R$, to the exterior schwarzschild spacetime given by the
following metric
\begin{eqnarray}
ds^{2}&=&-\left(1-\frac{2M}{r}\right)dt^{2}+\left(1-\frac{2M}{r}\right)^{-1}dr^{2} \nonumber\\
&&+r^{2}\left(d\theta^{2}+\sin^{2}\theta d\phi^{2}\right).
\end{eqnarray}
Here the matching occurs at a radius greater than the event
horizon which gives
\begin{equation}
\left(\frac{R}{b_0}\right)^{2}=1-\frac{2M}{R}~~~  and ~~~
1-\frac{b(R)}{R}=1-\frac{2M}{R},
\end{equation}
and using the Eq. (8) with the expression $e^{\nu(R)}=c_2^{2}R^{2}$, we
determine the values of the constants $c_2^{2}$,  $b_{0}$ and
total mass M as follows
\begin{eqnarray}
M&=&\frac{R}{\omega+3}-\frac{\tilde{\psi_0}}{2}R^{-\frac{3}{\omega}},\\
c_2^{2}&=&\frac{1}{R^{2}}\left[\frac{\omega+1}{\omega+3}+\tilde{\psi_0}R^{-\frac{3+\omega}{\omega}}\right],\\
b_0&=&\frac{R}{\frac{\omega+1}{\omega+3}+\tilde{\psi_0}R^{-\frac{3+\omega}{\omega}}}.
\end{eqnarray}

\section{TOV Equation}
An important step is to examine the stability of our present model under the different
forces namely gravitational, hydrostatics and anisotropic forces. This is simply by
considering the generalized Tolman-Oppenheimer-Volkov (TOV)
equation according to Ponce de Le$\acute{o}$n \cite{leon}
\begin{equation}
-\frac{M_G(r)(\rho+p_r)}{r}e^{\frac{\nu-\mu}{2}}-\frac{dp_r}{dr}+\frac{2}{r}(p_t-p_r)=0,
\end{equation}
where $M_G(r) $ represents the effective gravitational mass within the radius $r$,
which can derived from the Tolman-Whittaker formula and the explicit expression is
given by
\begin{equation}
M_G(r)=\frac{1}{2}re^{\frac{\mu-\nu}{2}\nu'}.
\end{equation}
Substituting the above expression in Eq. (26), we obtain the simple expression as
\begin{equation}
-\frac{\nu'}{2}(\rho+p_r)-\frac{dp_r}{dr}+\frac{2}{r}(p_t-p_r)=0.
\end{equation}

\begin{figure}[htbp]
\centering
 \includegraphics[scale=.4]{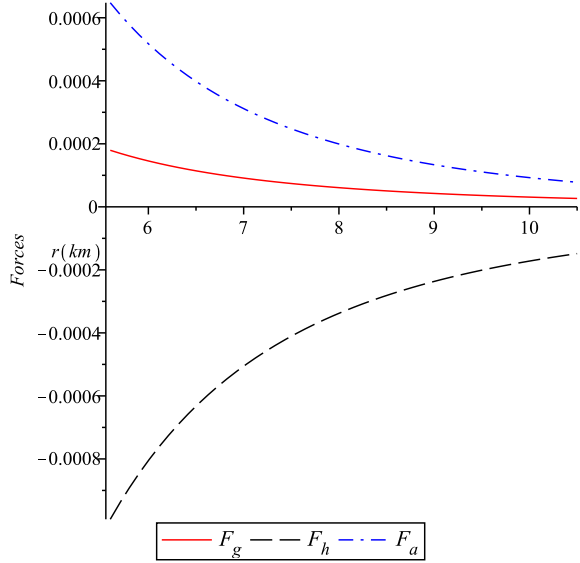}
 \caption{Variation of different forces acting on the wormhole are
 plotted against r with the same values of parameters as stated earlier in Fig. 1. }
  \label{fig:1}
\end{figure}
Therefore, one can write it in a more suitable form to generate the simpler equation
\begin{equation}
F_g+F_h+F_a=0,
\end{equation}
where $F_g =-\frac{\nu'}{2}(\rho+p_r)$, $F_h =-\frac{dp_r}{dr}$ and $F_a=\frac{2}{r}(p_t-p_r)$ represents the gravitational, hydrostatics and anisotropic forces, respectively.
Using the Eqs. (16-18), the above expression can be written as

\begin{eqnarray}
F_g=-\frac{(\omega+1)}{8\pi
r}\left[\frac{2}{r^2\left(\omega+3\right)}-\frac{\tilde{\psi_0}\left(2\omega+3\right)}
{\omega}r^{-\frac{3\left(\omega+1\right)}{\omega}}\right],\\
F_h=\frac{1}{8\pi}\left[\frac{4\omega}{r^3\left(\omega+3\right)}-\frac{3\tilde{\psi_0}\left(2\omega+3\right)\left(\omega+1\right)}{\omega
}r^{-\frac{\left(4\omega+3\right)}{\omega}}\right],\\
F_a=\frac{1}{4\pi
r}\left[\frac{1-\omega}{r^2\left(\omega+3\right)}+\frac{\tilde{\psi_0}\left(2\omega^2+3\omega-3\right)}
{3}r^{-\frac{3\left(\omega+1\right)}{\omega}}\right].
\end{eqnarray}
The profiles of $F_g, F_h,$ and $ F_a$ for our present model of wormhole are shown in Fig. $3$,
by assigning the same value of $\omega$ = -1.58 and $\tilde{\psi_0}= 0.09$ as we used in Fig.~$1$. It is clear
from the Fig. 3, that the hydrostatics force ($F_h$) is dominating compare to gravitational
($F_g$) and anisotropic forces ($F_a$), respectively. The interesting feature is that
$F_h$ takes the negative value while $F_g$ and $F_a$ are positive, which clearly
indicate that hydrostatics force is counterbalanced by the combine effect of gravitational
and anisotropic forces to hold the system in static equilibrium. There exist many excellent reviews on this
topic have been studied in-depth by Rahaman et al. \cite{rah16} and  Rani \& Jawad \cite{jawad}.

\begin{table}
\centering
\begin{tabular}{ |p{2cm}||p{2.5cm}|  }
 \hline
 \multicolumn{2}{|c|}{The obtained values of $E_g$  } \\
 \hline
 ~~~~~r  & ~~~~~~$E_g$ \\
 \hline
~~~~~6   &  ~~2.777766660\\
~~~~6.5 &  ~~2.711356872\\
~~~~~7   & ~~2.618496680\\
~~~~7.5 &  ~~2.506836836\\
~~~~~8   & ~~2.380794614\\
~~~~8.5 &  ~~2.243339402\\
~~~~~9   & ~~2.096641966\\
~~~~9.5 &  ~~1.942376899\\
~~~~10  &  ~~1.781885447\\
~~~10.45  &  ~~1.633036240\\
 \hline
\end{tabular}
\caption{The values of $E_g$ are obtained from the Eq. (37),
for the choices of the parameters $r_0^{+}=5.6$ km, $\tilde{\psi_0}=0.09$ and $\omega=-1.58$. }
\end{table}

\section{Active mass function and Total Gravitational Energy}

The active mass function for our wormhole ranging
from $r_0+$ ($r_0$ is the throat of the wormhole) up to the
radius R can be found as
\begin{equation}
M_{active}=\int_{r_0+}^{R}4\pi\rho r^{2}dr
=\left[\frac{r}{\omega+3}+\frac{\tilde{\psi_0}\left(2\omega+3\right)}{6}r^{\frac{-3}{\omega}}\right]_{r_0+}^{R}.
\end{equation}
The  active gravitational mass function of the wormhole is plotted in of Fig. 4 (left panel).
From the Fig. 4, we see that $M_{\mathrm{active}}$ is positive outside
the wormhole throat and monotonic increasing function of the radial co-ordinate, r.
\begin{figure*}[thbp]
\begin{center}
\begin{tabular}{rl}
\includegraphics[width=6 cm]{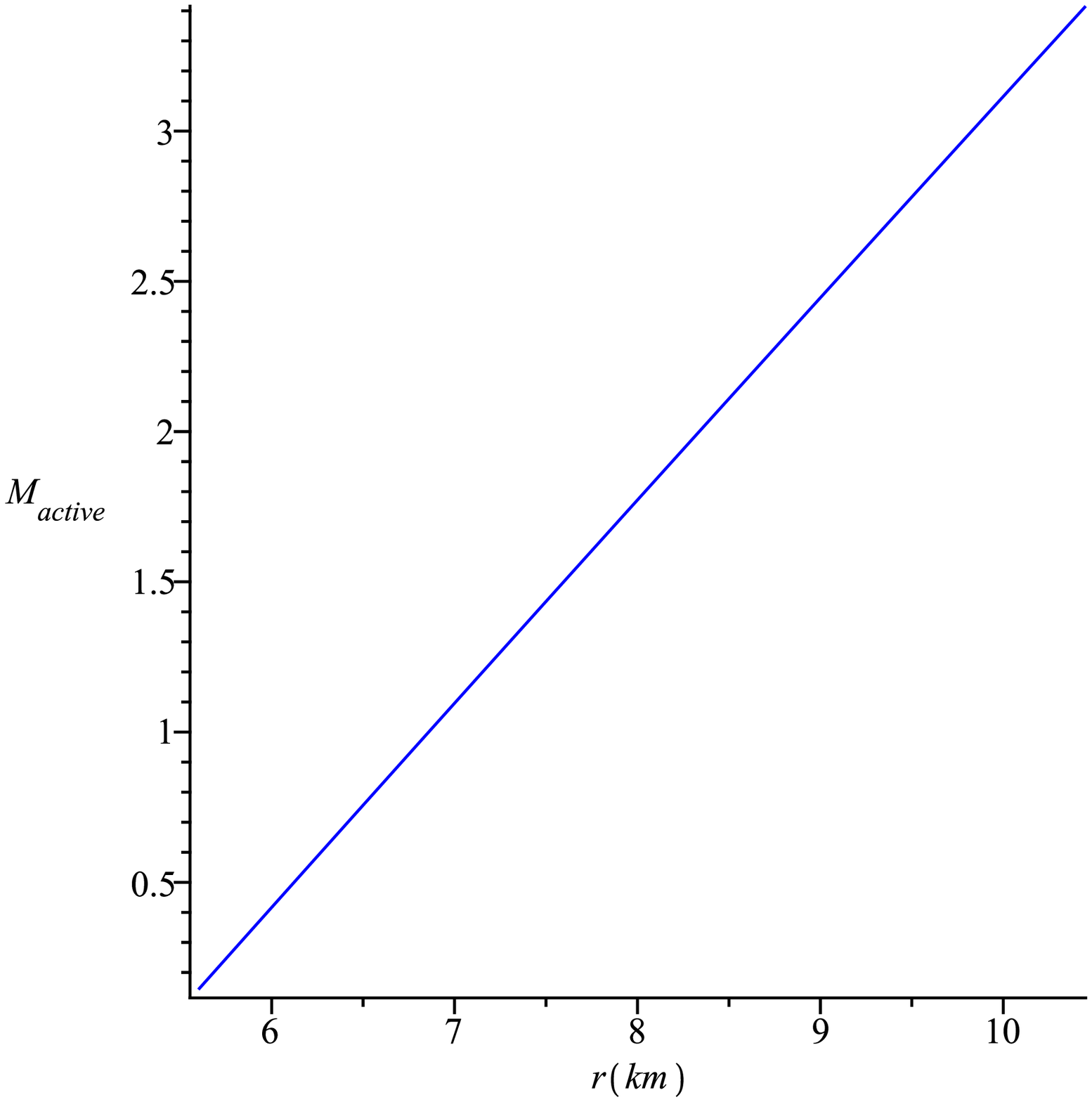}&
\includegraphics[width=6 cm]{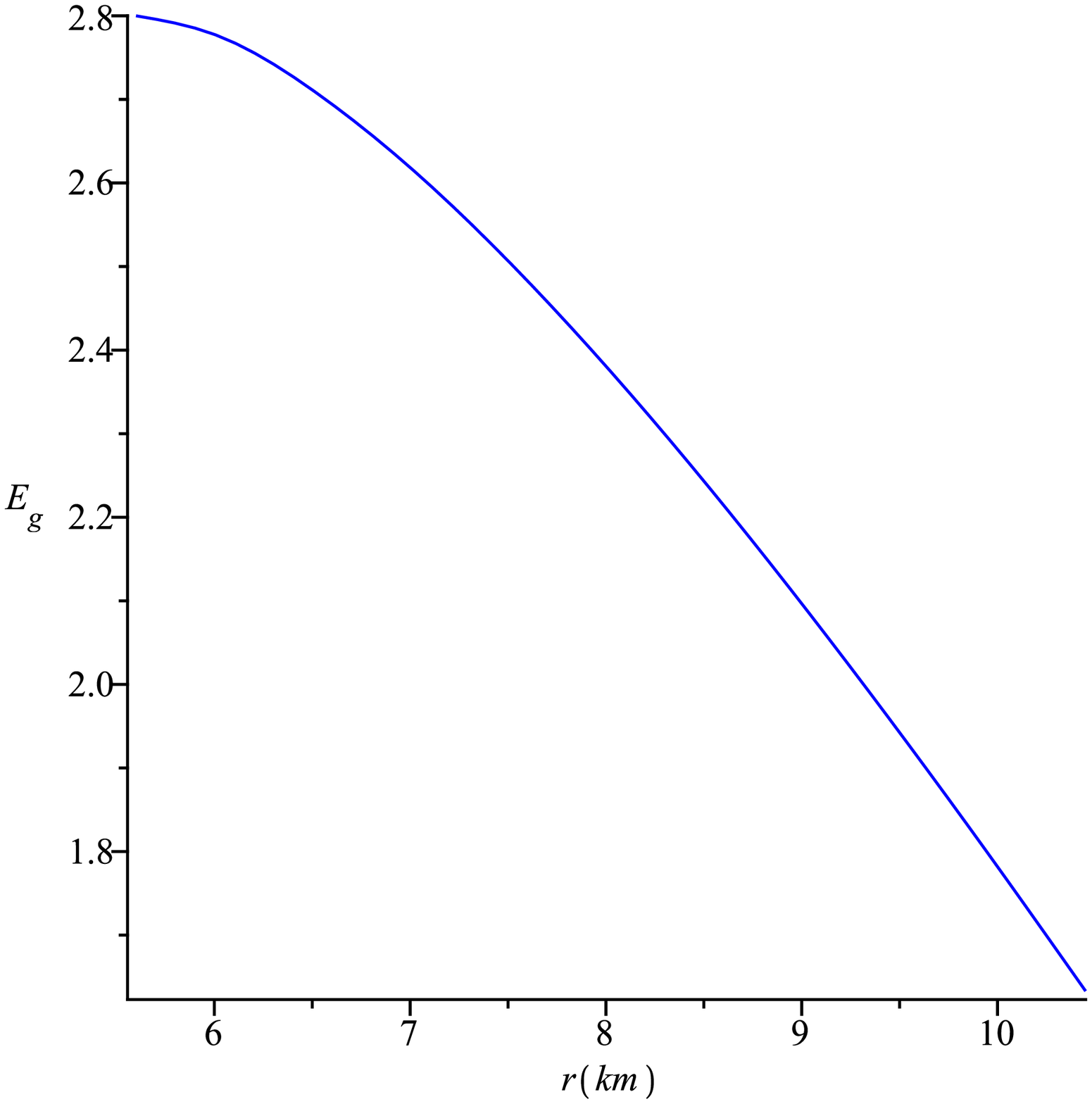}\\
\end{tabular}
\end{center}
\caption{Active gravitational mass of the wormhole is plotted against r in the left panel and Total gravitational energy is plotted against r in the right panel by taking the same values of the model parameters.}
\end{figure*}

For the study of total gravitational energy of the exotic matter inside a static wormhole
configuration we use the procedure adopted by Lyndell-Bell et al. and Nandi et al. \cite{lyn,Amrita,nandi}
for calculating the total gravitational energy $E_g$ of the wormhole, can be written in the form
\begin{equation}
E_g=Mc^{2}-E_M,
\end{equation}
where the total mass-energy within the region from the throat $r_0$  up to the
radius R can be provided as
\begin{equation}
Mc^{2}=\frac{1}{2}\int_{r_0^{+}}^{R}T_{0}^{0}r^{2}dr+\frac{r_0}{2},
\end{equation}
and the energy in other forms like kinetic energy, rest energy, internal energy etc. are defined by
 \begin{equation}
E_M=\frac{1}{2}\int_{r_0^{+}}^{R}\sqrt{g_{rr}}\rho r^{2}dr.
  \end{equation}
Note that here $\frac{4\pi}{8\pi}$ yields the factor $\frac{1}{2}$. By
taking into account Eqs. (34 - 36), we obtain
\begin{equation}
E_g=\frac{1}{2}\int_{r_0^{+}}^{R}[1-(g_{rr})^{\frac{1}{2}}]\rho r^{2} dr +\frac{r_0}{2},
 \end{equation}
where $g_{rr}=\left(1-\frac{b(r)}{r}\right)^{-1}$ and $r_0$ is the throat of the wormhole.
Now to find out the expression of total gravitational energy $E_g$, we have performed the integral of Eq. (37).
Due to the complexity of the coefficients $g_{rr}$ and $\rho$ we cannot extract analytical solution,
for that we solve the integral numerically. The numerical values of $E_g$
are obtained by taking $r_0^{+}=5.6$ Km. as a lower limit and by changing the upper limits,
which are given in Table. I.

\section{Gravitational Lensing}
We know that a photon follows a null geodesic $ds^2=0$, when external forces are absent.
Then the equation of motion of a photon can be  written as :
\begin{equation}
\dot{r}^2 +e^{-\lambda}r^2\dot{\phi}^2=e^{\nu-\lambda}c^2\dot{t}^2,
\end{equation}
where the dot represents derivative with respect to  the arbitrary affine parameter.
Since, neither $t$ nor $\phi$ appear explicitly in the variation principle,
their conjugate momenta yield the following constants of motion :
\begin{equation}
e^{\nu}c^2\dot{t}=E=\text{constant},
\end{equation}
\begin{equation}
r^2\dot{\phi}=L=\text{constant},
\end{equation}
where $ E$ and $ L$ are related with the conservation of energy and angular momentum, respectively.
Using these two constants of motion in Eq. (38), we get
\begin{equation}
\dot{r}^2+e^{-\lambda}\frac{L^2}{r^2}=\frac{E^2}{c^2}e^{-\nu-\lambda}.
\end{equation}

Now, using $r=\frac{1}{u}$ and eliminating the derivatives with respect to
the affine parameter by the help of  the conservation equations, we obtain
\begin{equation}\label{eq:sqdu}
\left(\frac{du}{d\phi}\right)^2+u^2=f(u).u^2+ \frac{1}{c^2}\frac{E^2}{L^2}e^{-\nu-\lambda}\equiv P(u),
\end{equation}
where $e^{-\lambda}=1-f(u)$, while from Eq. (19) yields
\begin{equation}
f(u)=\frac{2}{\omega+3}-\tilde\psi_0u^{\frac{\omega+3}{\omega}}.
\end{equation}
Moreover, from Eq. (42), we get
\begin{equation}
P(u)=\frac{u^2}{\omega+3}\left(2+\frac{E^2b_0^2(\omega+1)}{c^2L^2}\right)
+\tilde\psi_0u^{\frac{3(\omega+1)}{\omega}}\left(\frac{E^2b_0^2}{c^2L^2}-1\right).
\end{equation}
Let us proceed to discuss at the turning points \cite{lake}, the derivative of the
radial vector with respect to the affine parameter vanishes,
which in turn leads to $\frac{du}{d\phi}=0$. Consequently the turning point
is denoted by $r_\Sigma=1/u_\Sigma$ and given by
\begin{equation}\label{eq:turningpt}
r_\Sigma=\left(-\tilde{\psi_0}\frac{\omega+3}{\omega+1}\right)^{\frac{\omega}{\omega+3}}.
\end{equation}

\begin{table}
\centering
\begin{tabular}{ |p{2cm}||p{2.3cm}|p{2.5cm}|  }
 \hline
 \multicolumn{3}{|c|}{The deflection angles $\delta$ are listed for different values of w} \\
 \hline
 ~~~$\omega$  & ~~~~~~~$\phi$  & ~~~$\delta$ \\
 \hline
-1.38 &  1.126922 & -.887746 \\
-1.4  &   1.1130334357 & -.915523128\\
-1.58 &  1.0061484 & -1.1292932\\
-1.8  &  0.903820943 & -1.333948114\\
-2.1  &  0.7825234 & -1.5765432\\
-2.3  &  0.696393 & -1.748804\\
-2.4  &   0.98541384 & -1.17076232\\
-2.7  &  0.406932726 & -2.327724548\\
 \hline
\end{tabular}
\caption{The values of $\delta$ are obtained for the different choices of the
parameters $\omega$ when $\tilde{\psi_0}=0.09$. }
\end{table}

Differentiating Eq. (42) with respect to $\phi$, we get
\begin{equation}
\frac{d^2u}{d\phi^2}+u=Q(u),
\end{equation}
where $Q(u)=\frac{1}{2}\frac{dP(u)}{du}$, and define by
\begin{eqnarray}
Q(u)&=& \frac{u}{\omega+3}\left(2+\frac{E^2b_0^2(\omega+1)}{c^2L^2}\right) \nonumber \\
&\;& +\frac{3\tilde\psi_0(\omega+1)}{2\omega}u^{\frac{2\omega+3}{\omega}}\left(\frac{E^2b_0^2}{c^2L^2}-1\right).
\end{eqnarray}

Now, if the deflective source  are  absent, then the equation (46) modified to
\begin{equation}
u=\frac{cos(\phi)}{R},
\end{equation}
which is a straight line with R is the distance of closest approach to the wormhole.
This solution can treated as first approximation. Furthermore,
we use this solution as the first approximation to get the general solution.
This yields  the following form  of Eq. (47) as



\begin{equation}
\frac{d^2u}{d\phi^2}+u=A\cos\phi+B(\cos\phi)^{\frac{2\omega+3}{\omega}},
\end{equation}
where
{\small
\[A=\frac{2}{R(\omega+3)}+\frac{E^2b_0^2(\omega+1)}{Rc^2L^2(\omega+3)}, \mathrm{and} ~~
B=\frac{3\tilde\psi_0(\omega+1)}{2\omega}\left(\frac{E^2b_0^2}{c^2L^2}-1\right).\]}

With the aid of Eq. (49), the general solution is given by
\begin{eqnarray}\label{eq:trajectory}
u & = & \frac{\cos\phi}{R}+\frac{A}{2}\left(\cos\phi+\phi \sin\phi\right)+\nonumber\\
&\;& B\sin\phi\int\cos^{\frac{3(\omega+1)}{\omega}}\phi~~d\phi
+B\frac{\omega}{3(\omega+1)}\cos^{\frac{4\omega+3}{\omega}}\phi.\nonumber\\.
\end{eqnarray}
The light ray approaches from infinity at an asymptotic angle $\phi=-\left(\frac{\pi}{2}+\epsilon\right)$ and goes back to infinity at an asymptotic angle $\phi=\left(\frac{\pi}{2}+\epsilon\right)$. However the point of transition of the light ray from the Schwarzschild spacetime to the phantom spacetime is given by the turning points defined in (\ref{eq:turningpt}). The solution of the equation $~~u\left(\frac{\pi}{2}+\epsilon\right)=0$ yields the  angle $\epsilon$. The total deflection angle of the light ray can be obtained as  $\delta=2\epsilon$. In case of our wormhole, we have calculated  the  deflection angles for different values  of $\omega$ that  are tabulated in Table II. One can note that rather finding angle of deficit, we have found angle of surplus.

Lastly we must remember that the phantom spacetime under consideration, unlike  Schwarzschild, is basically non flat. Hence strictly speaking, an  asympotically straight line trajectory where $r\rightarrow\infty$  does not make sense. To resolve this issue we can consider the angles which the tangent to the light trajectory makes with the coordinate planes at a given point. Following Rindler and Ishak \cite{Rindler} we can find

\begin{equation}
\tan (\Psi)=\frac{r\left[e^{\nu(r)}\right]^{1/2}}{\left|\frac{dr}{d\phi}\right|}~,
\end{equation}
using Eq. [\ref{eq:sqdu}] we have
\begin{equation}
\tan(\Psi)=\sqrt{\frac{\frac{1}{R^2}\left[\frac{\omega+1}{\omega+3}+
\tilde\psi_0R^{-(3+\omega)/\omega}\right]}{\frac{AR-1}{r^2}+\frac{2B\omega}{3(\omega+1)}r^{-3(\omega+1)/\omega}}}.
\end{equation}

Assuming the bending angle to be small we can take $\tan(\Psi)\rightarrow\Psi$,
$\sin(\phi)\rightarrow\phi$ and $\cos(\phi)\rightarrow1$.
Thus from Eq. [\ref{eq:trajectory}] we can write the actual light deflection angle given by $|\epsilon|=\left|\Psi-\phi\right|$ as follows :

{\small
\begin{equation}
|\epsilon|=\left|\sqrt{\frac{\frac{1}{R^2}\left[\frac{\omega+1}{\omega+3}+\tilde\psi_0R^{-(3+\omega)/\omega}\right]}
{\frac{AR-1}{r^2}+\frac{2B\omega}{3(\omega+1)}r^{-3(\omega+1)/\omega}}}
-\left[\frac{\frac{1}{r}-\frac{1}{R}-\frac{A}{2}-\frac{B\omega}{3(\omega+1)}}{B+\frac{A}{2}}\right]\right|.
\end{equation}}

Working along the lines of Bhadra et. al. \cite{Bhadra} we can then calculate
the total deflection angle in terms of location of the source ($d_{LS}, \phi_S$)
and the observer ($d_{OL}, \phi_O$) as

{\footnotesize
\begin{eqnarray}
|\epsilon| &= & \left|\sqrt{\frac{\frac{1}{R^2}\left[\frac{\omega+1}{\omega+3}+\tilde\psi_0R^{-(3+\omega)/\omega}\right]}
{(AR-1)\left(\frac{1}{d_{LS}^2}+\frac{1}{d_{OL}^2}\right)+\frac{2B\omega}{3(\omega+1)}\left(d_{LS}^{-3(\omega+1)/\omega}
+d_{OL}^{-3(\omega+1)/\omega}\right)}} \right.\nonumber\\
&\;&\left.-\left[\frac{\frac{1}{d_{LS}}+\frac{1}{d_{OL}}-\frac{1}{R}-\frac{A}{2}-\frac{B\omega}
{3(\omega+1)}}{B+\frac{A}{2}}\right]~\right|\nonumber.\\
\end{eqnarray}}


\section{Junction Condition}
In previous section we matched our interior wormhole spacetime with the Schwarzschild exterior spacetime at the boundary $r=r_{\Sigma}$. Since the wormhole spacetime is not asymptotically flat we use the Darmois–Israel \cite{dm1,dm2} formation to determine the surface stresses at the junction boundary. The intrinsic surface stress energy tensor $S_{ij}$ is given by Lancozs equations in the following form
\begin{equation}
S^{i}_{j}=-\frac{1}{8\pi}(\kappa^{i}_j-\delta^{i}_j\kappa^{k}_k).
\end{equation}
The second fundamental form is presented by
\begin{equation}
K_{ij}^{\pm}=-n_{\nu}^{\pm}\left[\frac{\partial^{2}X_{\nu}}{\partial \xi^{i}\partial\xi^{j}}+\Gamma_{\alpha\beta}^{\nu}\frac{\partial X^{\alpha}}{\partial \xi^{i}}\frac{\partial X^{\beta}}{\partial \xi^{j}} \right]|_S,
\end{equation}
and the discontinuity in the second fundamental form is written as,
\begin{equation}
\kappa_{ij}=K_{ij}^{+}-K_{ij}^{-},
\end{equation}
where $n_{\nu}^{\pm}$ are the unit normal vectors defined by,
\begin{equation}
n_{\nu}^{\pm}=\pm\left|g^{\alpha\beta}\frac{\partial f}{\partial X^{\alpha}}\frac{\partial f}{\partial X^{\beta}}  \right|^{-\frac{1}{2}}\frac{\partial f}{\partial X^{\nu}},
\end{equation}
with $n^{\nu}n_{\nu}=1$. Where $\xi^{i}$ is the intrinsic coordinate on the shell. $+$ and $-$ corresponds to exterior i.e., Schwarzschild  spacetime and interior (our) spacetime respectively.\par

Considering the spherical symmetry of the spacetime surface stress energy tensor can be written as $S^{i}_j=diag(-\sigma,\mathcal{P})$. Where $\sigma$ and $\mathcal{P}$ is the surface energy density and surface pressure respectively.
The expression for surface energy density $\sigma$ and the surface pressure $\mathcal{P}$ at the junction surface $r = r_{\Sigma}$ are obtained as,
\begin{eqnarray}
\sigma &=&-\frac{1}{4\pi r_{\Sigma}}\left[\sqrt{e^{-\lambda}} \right]_{-}^{+} \nonumber\\
&&=-\frac{1}{4\pi r_{\Sigma}}\left[\sqrt{1-\frac{2M}{r_{\Sigma}}}-\sqrt{\frac{\omega+1}{\omega+3}+\tilde{\psi_0}r_{\Sigma}^{-\frac{3+\omega}{\omega}}}\right],
\end{eqnarray}

and

\begin{eqnarray}
\mathcal{P}&=&\frac{1}{8\pi r_{\Sigma}}\left[\left\{1+\frac{a\nu'}{2}\right\}\sqrt{e^{-\lambda}}\right]_{-}^{+}\nonumber\\
&&=\frac{1}{8\pi r_{\Sigma}}\left[\frac{1-\frac{M}{r_{\Sigma}}}{\sqrt{1-\frac{2M}{r_{\Sigma}}}}-\sqrt{\frac{\omega+1}{\omega+3}+\tilde{\psi_0}
r_{\Sigma}^{-\frac{3+\omega}{\omega}}}\right].
\end{eqnarray}

Hence we have matched our interior Wormhole solution to the exterior Schwarzschild spacetime in presence of thin shell.

\section{Discussion}

By the confirmation of various observational data that the Universe is undergoing a phase of
accelerated expansion and dark energy models have been proposed for this expansion. In the framework of
GR, the violation of NEC namely `exotic matter', is a fundamental ingredient of static traversable wormholes.
In this work, we investigated some of the characteristics
needed to support a traversable wormhole specific exotic form of dark energy, denoted phantom energy,
admitting conformal motion of Killing Vectors. We analyzed the physical properties and characteristics
of these wormholes with help of graphical representation. In the plots of Fig. 2 (left panel).
we obtain the throat of wormholes where $b(r) - r$ cuts the r-axis, is located at r = $r_0$ = 5.39 Km.,
which implies that $b(r)/r < 1$, met the flare-out condition. Another fundamental property of wormhole
is the violation of the null energy condition (NEC), also satisfy for our model given in Fig. 2 (right panel),
which provide a natural scenario for the existence of traversable wormholes. We discuss
the possibility of detection of wormholes  by means of gravitational lensing and we have found the angle of deflection to be negative i.e. angle of surplus. Additionally in order to give a physically feasible meaning to light deflection in a nonflat phantom spacetime, we have computed the  angles that the tangent to the light trajectory makes with the coordinate planes in terms of the location of the observer and the source.
Further we investigate, total gravitational energy content in the interior of exotic
matter distribution for the wormholes by Lyndell-Bell et al. \cite{lyn} and Nandi et al's.\cite{nandi} perception.
This lensing phenomena i.e. deflection of light by the wormhole   offers a good possibility to  detect the presence of wormhole. The present study offers a clue for possible detection of  wormholes   and  may encourage researchers  to seek observational evidence for wormholes.

\subsection*{Acknowledgments}
AB and FR would like to thank the authorities of the Inter-University Centre
for Astronomy and Astrophysics, Pune, India for providing research facilities. FR is also grateful to DST-SERB and DST-PURSE, Govt. of India for financial support. We are also  thankful to the referee for his constructive suggestions.

\end{document}